\documentclass[prb,twocolumn,superscriptaddress,floatfix,nopacs]{revtex4-2}
\usepackage{amsfonts,amsmath,amssymb,epsf,dcolumn,natbib,graphicx}
\usepackage{xcolor}
\usepackage{soul}
\usepackage{comment}
\usepackage[normalem]{ulem}
\usepackage{enumitem}
\usepackage[colorlinks=true, allcolors=blue]{hyperref}
\usepackage{hyperref}   

\newcommand{\beq}{\begin{equation}}
\newcommand{\eeq}{\end{equation}}
\newcommand{\be}{\begin{equation}}
\newcommand{\ee}{\end{equation}}
\newcommand{\bea}{\begin{eqnarray}}
\newcommand{\eea}{\end{eqnarray}}
\newcommand{\vvk}[1]{\textcolor{black}{#1}}  
\newcommand{\rev}[1]{\textcolor{black}{#1}}  
\newcommand{\AB}[1]{\textcolor{black}{#1}}
\newcommand{\balg}{\begin{align}}
\newcommand{\ealg}{\end{align}}

\begin{document}


\title{Direct simulations of H–He mixtures at planetary interior conditions: demixing, insulator–metal transition and miscibility boundaries}







\author{Valentin V.~Karasiev}
\email{Corresponding Author: vkarasev@lle.rochester.edu}
\affiliation{
Laboratory for Laser Energetics,
University of Rochester,
250 East River Road,
Rochester, New York 14623 USA
}

\author{S. X. Hu}
\affiliation{
Laboratory for Laser Energetics,
University of Rochester,
250 East River Road,
Rochester, New York 14623 USA
}
\affiliation{Department of Physics and Astronomy, University of Rochester, NY 14627, USA}
\affiliation{Department of Mechanical Engineering, University of Rochester, NY 14627, USA}

\author{Joshua P. Hinz}
\email{Present Address: Physics Department Southern University, 801 Harding Blvd, Baton Rouge, LA 70807, USA}
\affiliation{
Laboratory for Laser Energetics,
University of Rochester,
250 East River Road,
Rochester, New York 14623 USA;}

\author{R. M. N. Goshadze}
\affiliation{
Laboratory for Laser Energetics,
University of Rochester,
250 East River Road,
Rochester, New York 14623 USA
}

\author{Shuai Zhang}
\affiliation{
Laboratory for Laser Energetics,
University of Rochester,
250 East River Road,
Rochester, New York 14623 USA
}

\author{Armin Bergermann}
\affiliation{
Institut für Physik, Universität Rostock, D-18051 Rostock, Germany
}

\author{Ronald Redmer}
\affiliation{
Institut für Physik, Universität Rostock, D-18051 Rostock, Germany
}

\date{May 24, 2026}

\begin{abstract}
Accurate knowledge of the electrical and thermal conductivities and structural properties of hydrogen-helium mixtures under thermodynamic conditions within and beyond the immiscibility range is very important to predict the thermal evolution and internal structure of gas giant planets like Jupiter and Saturn. Here, we propose a novel method to determine the immiscibility boundary accurately 
\vvk{without the need for free energy calculations, while providing consistent insights into structural and transport properties of mixtures}.
We show with direct large-scale {\it ab initio} simulations that the insulator–metal transition (IMT) of the hydrogen subsystem is strongly affected by an admixture with a small fraction of helium and occurs at temperatures significantly higher than those of pure hydrogen. At pressures below 150~GPa, \vvk{the IMT boundary is not related anymore to the H$_2$ subsystem dissociation,} the system remains insulating even after the full dissociation of H$_2$ molecules and its transition to an H-He mixture. 
The offset of the IMT in the H–He mixture relative to the dissociation region in the hydrogen subsystem and the significant reduction of static electrical and thermal conductivity by a factor between two and a few thousand relative to pure hydrogen found in mixtures have consequences for Jupiter and Saturn's thermal evolution, internal structure, and dynamo action, affecting a large fraction of the interior of both planets. 
\end{abstract}
                        
\maketitle


Hydrogen (H) and Helium (He) are the most abundant elements in the universe and the main constituents of gas giant planets such as Saturn, Jupiter, and extrasolar planets~\cite{1984ApJ...282..807C,https://doi.org/10.1029/98JE00695, Helled2020b}. The reliability of planetary models predicting the interior structure and the dynamo generation strongly depends on the accuracy of predictions for the material properties; see, e.g.~\cite{Helled2020b, Nettelmann2012, Nettelmann2008, Nettelmann2013a, Militzer2021, PhysRevB.87.014202, Mankovich_2020, guillot:hal-00991246, Walsh2011, Batygin2015}. {\it Ab initio} molecular dynamics (AIMD) simulations based on the free energy density functional theory (DFT) have proven to be a successful and key tool to investigate materials by calculating thermodynamic and transport properties and predicting phase transition regions at extreme pressure and temperature conditions relevant for planetary interiors~\cite{Graziani2014}. In addition to the equation-of-state (EOS) data, the structural and transport properties (e.g., the electrical and thermal conductivity) and the H-He demixing region are crucial for modeling gas giant planets. The demixing of H-He leads to the formation of He droplets that rain towards the planetary core, thereby increasing the planet's internal heat budget~\cite{Howard2024, Puestow2016, Stevenson1977b, Fortney2004, Mankovich_2020}.

The most successful approach to calculate the H-He miscibility gap is to evaluate differences in the Gibbs free energy of mixing $(\Delta G)$ based on the non-ideal entropy of mixing~\cite{PhysRevLett.120.115703,doi:10.1073/pnas.0812581106}. \AB{Such differences in $\Delta G$ demonstrate that the demixed state is lower in free energy than the mixed state, and thus constitute the fundamental thermodynamic reason why H and He remain immiscible under the relevant pressure–temperature conditions.}
Approaches based on $\Delta G$ consist of two steps: (i) AIMD simulations for a small system with only a few particles to establish EOS tables for a perfectly mixed system; and (ii) calculation of the non-ideal entropy via a combination of coupling constant integration (CCI) and thermodynamic integration (TDI) over the established EOS. This approach successfully establishes the H-He miscibility diagram but without any insight into structural and transport properties. These calculations use relatively small system (e.g. 64 electrons~\cite{PhysRevLett.120.115703}) representing a compromise to avoid demixing effects already in the simulation cell and to obtain reasonably well-converged thermodynamic data. However, the question remains whether this method produces converged and reliable results. 

Direct evidence for H-He demixing was found by performing large-scale simulations \cite{PhysRevB.84.235109,PhysRevB.84.165110,PhysRevB.87.014202}. In these calculations, the regions rich and poor in He were identified by visual observation of typical MD snapshots under conditions where H and He are nearly completely immiscible. Note that the authors had not been able to examine the precise $P$-$T$ conditions of the H-He immiscibility boundary. Some qualitative features of the radial distribution function (RDF) for H-H, He-He, and H-He, indicating demixing, were very pronounced for conditions that are deep in the immiscibility region. 

Chang~{\it et al.}~\cite{Chang.NC2024} used AIMD data to train a neural network potential (NNP) and conducted 
large super-cell MD simulations overcoming finite-size effects. The H-He miscibility gap was quantified based on reweighted conditional probability \lq\lq atoms in neighborhoods". This work predicts a significantly higher mixing temperature compared to the results based on evaluating differences in $\Delta G$. Note that an earlier experimental campaign suggested the miscibility gap at even higher temperatures~\cite{Brygoo2021}.

In this Letter, we develop a novel approach investigating the demixing boundary together with investigating structural properties of H-He mixtures by large supercell AIMD simulations, thus remedying the major deficiency of the $\Delta G$ based methods -- a lack of consistent insights into structural and transport properties of mixtures. We found that a drop of the first peak in the H-He RDF along an isobar provides a sharp \lq\lq mechanical" signature of demixing. Therefore, this behavior allows us to determine the size of the miscibility gap without any further analysis. This simulation approach not only leads to well-converged results, but also maintains the accuracy and advantages of AIMD simulations avoiding the inherent shortcomings of NNPs that simulate liquid H \cite{Karasiev.Nature2021,Cheng2021,PhysRevE.111.045307}. 
\rev{The approach developed in the present work has been used by few of our co-authors to investigate immiscibility boundary and transport properties in Ne-H mixtures \cite{bergermann2026miscibilitytransportpropertieshydrogenneon}, that demonstrates applicability of the approach to other mixtures at wide range of mixing parameters.}

%
\begin{figure}
\includegraphics[angle=-00,height=3.6cm]{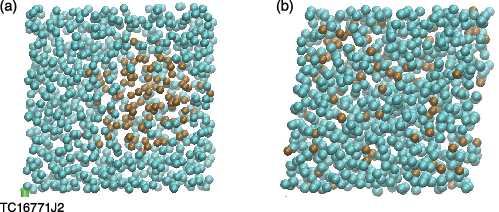}
\caption{
\vvk{An example of H$_2$-He demixing (a) and H-He mixing (b) occurring directly in the simulation box 
for the He$_{104}$H$_{816}$ system (x=0.11304) at $P=150$ GPa, and $T=1000$ and 6500 K respectively
(He and H atoms are shown as gold and cyan spheres respectively).}
}
\label{Fig1}
\end{figure}
%

\textsf{\textbf{AIMD simulations.}}
We performed large supercell AIMD simulations in the isothermal-isobaric ($NPT$) ensemble. Thermal XC effects,  which are not negligible at some $P$-$T$ conditions considered in this work, were taken into account using the Karasiev-Dufty-Trickey (KDT16) GGA level functional \cite{KDT2018PRL}. 
\vvk{This choice of functional is justified by KDT16 being the free-energy  counterpart of the oft-used ground-state PBE functional.} 
Simulations were performed for systems with $1024$ electrons for two helium fractions, ${\mathrm x}=0.11304$ (He$_{104}$H$_{816}$) and ${\mathrm x}=0.27522$ (He$_{221}$H$_{582}$)\vvk{, defined as x$\equiv N_{\rm He}/(N_{\rm H}+N_{\rm He})$.}
%
%

\rev{We used a plane wave cutoff of $1400$~eV. In all simulations, the Baldereschi mean-value $k$-point (BMVP)~\cite{PhysRevB.7.5212} was used to sample the Brillouin zone of (near-)cubic supercells. The number of thermally occupied bands included in our simulations was large enough to ensure the highest energy state is occupied around $0.2 \times 10^{-5}$ or below. The MD time step, depending on temperature, varied between $0.2$ and $1.0$~fs. After reaching equilibrium (usually $5000$ MD steps), we run our simulations for up to $40,000$ MD steps. We followed Ref.~\cite{Karasiev.Nature2021} for AIMD simulations of pure H.}
\rev{When we start simulations from a perfectly mixed configuration at conditions when the system is expected to be demixed, the system demixes within ~2.5-5.0 ps (though starting from an artificially formed \lq\lq demixed" configuration accelerates the equilibration process). Demixing in many cases can be observed by a visual inspection of simulation cell snapshots. We keep running AIMD for up to 30 ps, and the system remains in the same demixed phase. This clearly demonstrates the convergence with respect to the AIMD time length.}

We calculated density profiles, direct current (dc), and thermal conductivities of H-He mixtures along selected isobars to study the miscibility gap, structural properties, the dissociation transition from H$_2$-He to H-He, and the corresponding insulator-to-metal transition.

Electrical and thermal conductivities were calculated using the Kubo-Greenwood~\cite{Kubo1957, Greenwood1958} formalism implemented in the {\sc VASP} \cite{kresse1996efficient} and {\sc KGEC} \cite{CALDERIN2017118} packages. The insulator-metal transition  (IMT) has been determined according to Mott's criterion for the minimum metallic conductivity, extended to finite temperatures~\cite{Edwards2010, PhysRevB.84.235109}, with values of $2000${\raisebox{0.5ex}{\tiny$\substack{+3000 \\ -1000}$}}~S/cm. The typical number of statistically independent snapshots used for calculations along AIMD trajectories was between 100 and 150 for the mixtures. In contrast, we used $21$ snapshots for pure H. 
Calculations used the Baldereschi mean-value $k$-point ~\cite{PhysRevB.7.5212}.
To have a more accurate characterization of the insulator-metal transition than possible from a GGA, we used the Heyd-Scuseria-Ernzerhof XC functional \cite{10.1063/1.1564060,1Heyd_erratum_2006} for selected $P$-$T$ conditions.

\textsf{\textbf{A novel approach for miscibility gap predictions.}}
For $P$-$T$ conditions under which H and He are immiscible, our 
AIMD 
simulations exhibit direct evidence of demixing, with He-rich and He-poor regions clearly identifiable within the simulation cell.
%
It is clear that the H-He interface in a two-phase demixed state\vvk{, (when He droplets are formed),} is a surface. Hence, the probability for a He atom to find an H atom at low distances is reduced compared to the perfectly mixed system when the H-He interface is \lq\lq volumetric"\vvk{, meaning that H-He neighbors are distributed near-uniformly through the volume}. \vvk{Figure \ref{Fig1} shows an example of representative snapshots in the demixed (left panel) and mixed (right panel) states.} These probability properties are reflected in the magnitude of the first peak in the H-He RDF. 
The magnitude of this peak along an isobar provides a sharp quantitative signature of the H-He demixing and its transition to a perfectly mixed state. In sufficiently small H-He systems, demixing is avoided under all thermodynamic conditions~\cite{PhysRevLett.120.115703}. Accordingly, the H-He \lq\lq volumetric" (or bulk) interface enhances the probability that a He atom is close to an H atom. 
In contrast, in large, demixed H-He systems, the first peak in the H-He RDF is expected to be significantly lower compared to a small, mixed system, as the bulk H-He interface reduces to a surface. The H-He demixing is most pronounced at the lowest temperatures but becomes less pronounced at higher temperatures until H-He is completely miscible. Accordingly, increasing the system's temperature increases the first peak in the H-He RDF. Subsequently, the first H-He RDF peak reaches the maximum upon perfect mixing and is expected to decrease with further temperature increase \vvk{(due to thermal expansion)}. Such a simple analysis of the H-He RDF behavior along isobars provides a single quantity that accurately and efficiently characterizes the demixing state independently of the shape of the He-rich regions. 

\begin{figure}
\includegraphics[angle=-00,height=5.0cm]{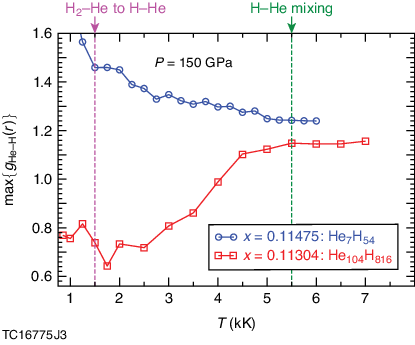}
\caption{Magnitude of the first H-He RDF peak as a function of temperature for pressure of $150$ \rev{GPa} 
\vvk{shown for a small always mixed system (blue curve), and for large one that transits from demixed to mixed state (red curve)}. The vertical green dashed line indicates the transition to the perfectly mixed state of liquid H-He (He$_{104}$H$_{816}$, ${\mathrm x}=0.11304$) mixtures. Vertical magenta dashed lines depict the H$_2$-He to H-He transition.}
\label{RDFs1}
\end{figure}
%

\textsf{\textbf{Miscibility Gap.}}
Figure~\ref{RDFs1} shows the magnitude of the first peaks of the H-He RDF as a function of temperature along the pressure of $150$ 
and He fraction ${\mathrm x}=0.11304$ \vvk{for large (red curve) and small (x=0.11475, blue curve) systems}. 
The H-He system transfers from a demixed to a mixed state once the curve reaches its maximum approaching the mixed small system value (vertical green dashed line). 
At the lowest temperature of $850$~K, the system is a mixture of molecular hydrogen and helium. 
Note that the magnitude of the first peak in the H-He RDF for a demixed system with He$_{104}$H$_{816}$ atoms is significantly smaller ($\approx$ 0.77) compared to the value for the perfectly mixed system with only He$_7$H$_{54}$ atoms, 
which is $\approx$ 1.92. Increasing the temperature 
leads to the dissociation of H$_2$ molecules. 
The magnitude of the first peaks slightly increases at a temperature of $1250$~K. \rev{It drops to $0.64$ upon the H$_2$ subsystem's dissociation at $1500$~k, which at this pressure is accompanied by the metallization of the mixture
(see Fig. \ref{RHO-SIGMA1} below). 
This result confirms that the metallization of the H$_2$ subsystem is not the primary driver, albeit it fosters the H-He demixing process.}

With a further temperature increase, the magnitude of the first H-He RDF peak increases, reaching its maximum at $5500$~K and becoming essentially flat at even higher temperatures. 
\vvk{Note that there are two competing processes associated with the temperature increase. One is the diminution of demixing (i.e. partial mixing) which tends to increase the magnitude of the first peak. In contrast, thermal expansion, the other process, decreases the magnitude.}
Subsequently, the maximum reached at $5500$~K corresponds to the transition to a near-perfectly mixed state. We assign an error (an upper bound estimation) of $\pm$ 500 K to this miscibility boundary temperature. 
Results for other $P$-$T$ conditions and for one more value of the He fraction can be found in the SI.


%
\begin{figure}
\includegraphics[angle=-00,height=5.5cm]{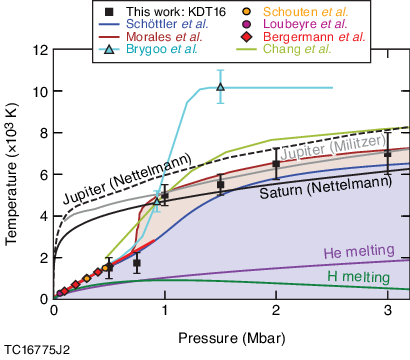}
\caption{Miscibility diagram for solar He abundance. Our results using the KDT16 functional are shown by black squares. Our prediction is compared to earlier theoretical results: Schöttler and Redmer~\cite{Schoettler2018} (blue line), Morales {\it et al.}~\cite{doi:10.1073/pnas.0812581106} (brown curve), Schouten {\it et al.}~\cite{Schouten1991} (orange circles), Bergermann {\it et al.}~\cite{Bergermann2021a} (red diamonds), and Chang {\it et al.}~\cite{Chang.NC2024}. Experimental results are shown as colored symbols: Loubeyre {\it et al.}~\cite{Loubeyre1987} (purple circles) and Brygoo {\it et al.}~\cite{Brygoo2021} (cyan triangles). Present day planetary isentropes for Jupiter are shown as a black dashed curve~\cite{Nettelmann2008} and solid gray curve \cite{Hubbard2016}, while the isentrope for Saturn is depicted by a solid black curve~\cite{Nettelmann2013a}.}
\label{Fig5}
\end{figure}

%
Figure~\ref{Fig5} presents the miscibility diagram for solar He mass concentration $Y=0.28$ with our results depicted as black squares. The H-He immiscibility boundary, as identified by our simulations using the thermal KDT16 XC functional, is approximately \rev{$500 \pm 250$} K higher than the ground state PBE \cite{PBE1996} results (see comparisons shown in Fig. S5 in the SI). 
That \rev{$+500\pm 250$} K offset arises from thermal XC contributions in KDT16. It is very close to the $+540$~K additive offset in the evolution model of Saturn by Mankovich and Fortney\cite{Mankovich_2020,Preising_2023}.

Our large-scale AIMD 
simulations are in qualitative agreement with earlier simulations~\cite{Schoettler2018, Bergermann2021a, doi:10.1073/pnas.0812581106} even though they might suffer from finite-size effects due to shorter simulation times and smaller particle numbers. In contrast, the results of Chang {\it et al.}~\cite{Chang.NC2024},  based on MD simulations driven by an NNP, indicate H-He demixing at temperatures approximately $1000$~K higher. Note that we cannot confirm the significantly higher temperatures found experimentally by Brygoo {\it et al.}~\cite{Brygoo2021}, although we used a thermal GGA XC functional and conducted large-scale MD simulations. However, our 
findings agree well with recent interior and evolution models of Saturn and Jupiter~\cite{Mankovich_2020, Howard2024}.

\textsf{\textbf{Structural properties, and static electrical and thermal conductivity.}}
The first-order liquid-liquid phase transition in pure dense H coincides with the dissociation of the H$_2$ molecules. Transition signatures include density jumps along isobars 
and corresponding jumps in dc conductivity, as well as sharp changes in H-H RDFs~\cite{Karasiev.Nature2021, Bonitz2024arXiv, Bergermann2024a, Hinz2020, Scandolo2003, Geng2019, Morales2010a, Lorenzen2010, Pierleoni2016, Mazzola2018}. 


Density profiles and dc conductivities for pressures of \vvk{$150$ and} $75$ GPa, and 
for He fractions of ${\mathrm{x}}=0.0$, ${\mathrm{x}}=0.11304$, 
are presented in Figs.~\vvk{\ref{RHO-SIGMA1}} and \ref{Fig2}, where $\mathrm{x}=0.0$ corresponds to the pure H system. \rev{The dissociation of pure molecular H$_2$, accompanied by an increase in density and dc conductivity, leads to metallization}. This increase is consistent with a drop of the first peak of the H-H RDF, indicating the dissociation process (see Fig.~S11  in the SI). However, a small amount of He strongly affects the dissociation and metallization processes of H: (i) the density increase in the mixture becomes much smoother or converts to a plateau depending on the He-fraction and thermodynamic conditions; (ii) the dissociation process in the mixtures is delayed by $\approx$ 250~K (or more depending on He fraction and pressure) as a consequence of the H$_2$ molecule bond strengthening~\cite{PhysRevB.75.024206}; (iii) the metallization temperature of the H$_2$-He mixture, \rev{determined according to Mott’s criterion for the minimum metallic conductivity of 2000 S/cm, at $P=150$ GPa coincides with the H$_2$ subsystem dissociation. The IMT temperature of 1500 K is right in the middle of the H$_2$ subsystem dissociation defined by the temperature range between 1250 K and 1750 K), see Fig. \ref{RHO-SIGMA1}}. However, at $P=75$ GPa the metallization temperature (4000 K) is significantly higher compared to the temperature of dissociation of the H$_2$ subsystem (2750 K) (see Fig. \ref{Fig2}), leading to an offset of IMT relative to the dissociation of the H$_2$ subsystem in a mixture. As a consequence of this, mixtures of atomic H and He remain insulating in a wide range of thermodynamic conditions. 
Figure \ref{Fig3} depicts the different $P$-$T$ conditions for the IMT of an H-He mixture compared to the dissociation in the H subsystem for two fractions x. The shaded areas indicate an insulating mixture of He and atomic H.

\begin{figure}
\includegraphics[angle=-00,height=5.5cm]{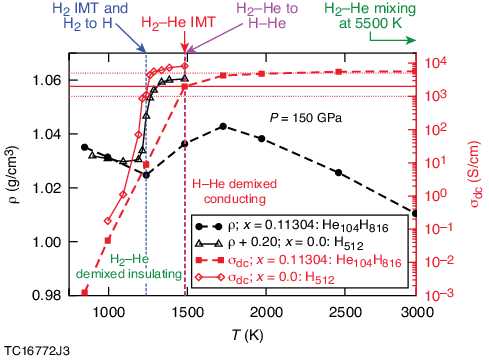}
\caption{\vvk{Density and dc conductivity of H-He mixtures as compared to pure H along  the $P=$ 150 isobar indicating temperatures of the H$_2$-He to H-He, IMT and H-He demixing/mixing transitions in liquid He$_{104}$H$_{816}$ (x=0.11304).
The IMT of the H$_2$-He mixture coincides with the H$_2$ subsystem dissociation and occurs at $T=1500$ K. The density of pure H (x=0.0) is shifted by 0.20 g/cm$^3$ for better visualization. 
}}
\label{RHO-SIGMA1}
\end{figure}

\begin{figure}
\includegraphics[angle=-00,height=5.5cm]{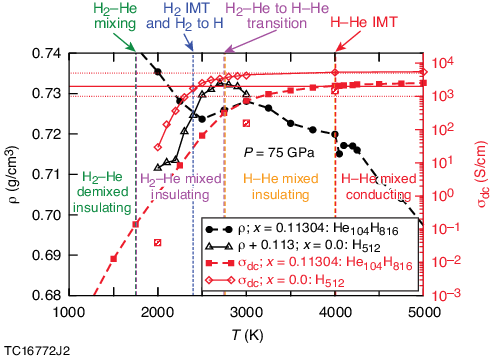}
\caption{
Density and dc conductivity of H-He mixtures as compared to pure H along  the $P=$ 75 isobar indicating temperatures of the H$_2$-He to H-He, IMT and H-He demixing/mixing transitions in liquid He$_{104}$H$_{816}$ (x=0.11304). 
\vvk{The density of pure H (x=0.0) is shifted by 0.113 g/cm$^3$ for better visualization. With the temperature increase, the H-He system remains insulating even after the H$_2$ subsystem dissociation.}
Three red squares filled with diagonal pattern correspond to $\sigma_{\rm dc}$ values calculated with hybrid HSE06 functional.
}
\label{Fig2}
\end{figure}

Kohn-Sham band gaps from most semi-local XC functionals greatly underestimate the fundamental gap~\cite{PBE1996}.
Conductivities, especially before the transition to a metallic state, are overestimated as a consequence. Hybrid functionals, such as HSE06~\cite{1Heyd_erratum_2006} for example, provide more realistic values for band gaps and conductivities. To estimate the reliability of the metallization temperature predicted by the semi-local thermal KDT16, we performed Kubo-Greenwood reference calculations with the HSE06 hybrid for $P=75$ GPa at three selected temperatures. HSE06 results, shown in Fig. \ref{Fig2}, clearly indicate that KDT16 predicts dc conductivity reasonably well near the metallization temperature $T=4000$ K; however, conductivity values are significantly overestimated in the insulating regime. Thus, the IMT 
temperatures, as predicted in our work and summarized in Fig. \ref{Fig3}, can be considered as a lower bound, i.e. in reality the metallization 
temperature might be even higher by a few hundreds of Kelvin.
A small kink in the density profile observed at a temperature of $4000$ K and a pressure of $75$~GPa (Fig. \ref{Fig2}), surprisingly coincides with the IMT, most probably a consequence of pressure ionisation as explained in an earlier work~\cite{Preising_2023}. 
%
%

%
\begin{figure}
\includegraphics[angle=-00,height=3.7cm]{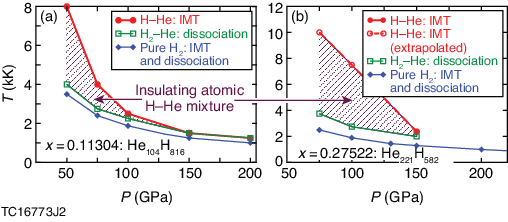}
\caption{
	The offset of insulator-metal transition relative to the H subsystem  dissociation in H-He mixtures.
	(a) H-He mixture insulator-metal transition and H$_2$-He $\rightarrow$ H-He dissociation  
	curves for the He$_{104}$H$_{816}$ (x=0.11304) mixture are compared to the IMT and dissociation boundary of pure H$_2$.
	(b) Same as in (a) for He$_{221}$H$_{582}$ (x=0.27522) mixture. \vvk{The IMT boundary at $P=75$ GPa (open red circle), was estimated by extrapolation of calculated data at 150 and 100 GPa.} Both panels show a wide range of thermodynamic conditions
	when atomic H - He mixture is insulating.
}
\label{Fig3}
\end{figure}
%

\begin{figure}
\includegraphics[angle=-00,height=4.4cm]{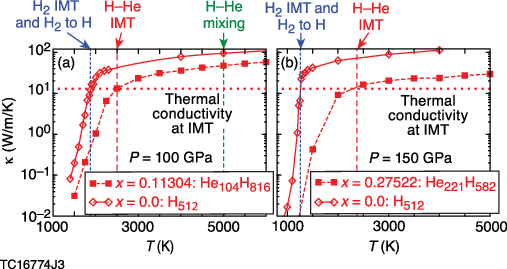}
\caption{
	Electronic thermal conductivity of H-He mixtures as predicted by DFT simulations. (a) Liquid He$_{104}$H$_{816}$ mixture (x=0.11304) along the $P=100$ GPa. (b) Liquid He$_{221}$H$_{582}$ mixture (x=0.27522) \vvk{along the $P=150$ GPa isobar}. Comparisons are made to the pure H system. Horizontal dashed line shows a near-universal value ($\approx 13$ W/m/K) of the electronic thermal conductivity reached upon the insulator-metal transition for each system.
}
\label{Fig4}
\end{figure}

In Fig.~\ref{Fig4}, we depict thermal conductivities along the $100$ and $150$~GPa isobars for ${\mathrm x}=0.0$, ${\mathrm x}=0.11304$, and ${\mathrm x}=0.27522$. The thermal conductivities in mixtures are generally lower compared to pure H, and are approximately two to three orders of magnitude lower at temperatures corresponding to the IMT in pure H. These results can be expected considering that thermal conductivities rise with increasing electrical conductivity and given the significant offset of H-He mixture metallization relative to the pure molecular H$_2$ system (see Fig. \ref{Fig2}). Note that the thermal conductivity of pure H and H-He mixtures reaches a nearly universal value of $\approx 13$~W/m/K (indicated by a horizontal dashed line) around the IMT defined according to the nominal value of the Mott criterion, which is expected to distinguish between the metallic and non-metallic behavior for relatively low transition temperatures. 

The corresponding value of the thermal conductivity of approximately $13$~W/m/K is an analog of the Mott minimum metallic electrical conductivity, distinguishing between the metallic and non-metallic phases in terms of thermal conductivity. This minimum metallic thermal conductivity criterion requires a semi-empirical extension to elevated temperatures, similar to the discussion of Mott's minimum metallic electrical conductivity.~\cite{Edwards2010, PhysRevB.87.014202}

\begin{figure}
\centering
\includegraphics[angle=-00,height=5.0cm]{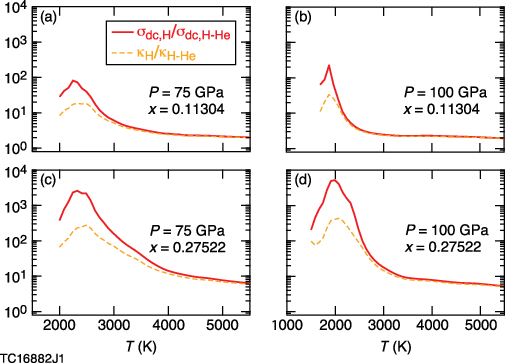}
	\caption{
	The ratio of pure hydrogen and H-He mixture conductivities (dc -- red solid, and thermal -- dashed orange curves)
	along selected isobars for He fractions x$=0.11304$ (panel (a) along the $P=75$ GPa isobar, and (b) along the $P=100$ GPa isobar) and x$=0.27522$ (panel (c) along the $P=75$ GPa isobar, and (d) along the $P=100$ GPa isobar). 
}
\label{Fig6}
\end{figure}

Figure~\ref{Fig6} delineates the drastic reduction in electrical and thermal conductivities in H-He compared to the pure H system for isobars of $75$ and $100$~GPa. This reduction reaches its maximum at $P$-$T$ conditions close to the molecular to atomic transition in pure H. An increase of the He fraction from x$=0.11304$ to x$=0.27522$ further decreases the conductivities yielding differences of a few orders in magnitude.



The enormous increase in computational power \vvk{during past decade} enabled us to combine the best of two worlds: large-scale MD simulations with the accuracy of \textit{ab inito} simulations without training an NNP as an intermediate step. Therefore, our results reliably predict the $P$-$T$ conditions of the miscibility gap without finite-size effects or the additional approximations of a data-driven approach; see our convergence tests in the SI. \AB{Additionally, our method avoids the cumbersome calculation of the non-ideal entropy necessary to evaluate differences in the free enthalpy~\cite{Schoettler2018, doi:10.1073/pnas.0812581106}}. An admixture with a small atomic fraction of He (
x=$0.11304$) reduces the conductivity by a factor between $\approx$ 2 and several tens. Increasing the He fraction further decreases the conductivity and yields a difference of a few orders in magnitudes. This effect had been investigated only qualitatively in earlier works~\cite{Lorenzen2010, PhysRevB.75.024206}. For example, at pressures below $150$~GPa, the H-He system remains insulating even after the H$_2$ molecular subsystem is completely dissociated. This result has important implications not only for modeling the interiors of Jupiter and Saturn -- particularly regarding the interplay between thermal conduction and convection in the demixing zone -- but also for our understanding of dynamo formation. The thermal conductivity strongly affects the thickness and stability of the H-He demixing layer and its corresponding non-adiabatic $P$-$T$ profile~\cite{Markham2024}.

It has been suggested that the metallization of the H-He system acts as a catalyst for H-He demixing \cite{PhysRevB.84.235109, Knudson2015}. Such a connection between the system metallization and the immiscibility boundary is not consistent with the non-ideal entropy miscibility diagram. In our simulations, we observed that the insulating mixture of molecular hydrogen and helium (H$_2$-He) spontaneously separates into H$_2$- and He-rich phases. \rev{Nevertheless, metallization of the H$_2$ subsystem, if occurs within the demixed regime, significantly enhances the demixing process.}
\\

\rev{
\textsf{\textbf{Supporting Information.}}
The Supporting Information is available free of charge at http://.... . 
\\
Additional computational details, convergence tests, and additional results on the immiscibility boundary and on the structural and conducting properties.
}
\\

\textsf{\textbf{Acknowledgments.}}
V.V.K. acknowledges with thanks informative conversations with Sam Trickey.
VVK, SXH, JPH, RMNG, and SZ acknowledge the support by the Department of Energy [National Nuclear Security Administration] University of Rochester ``National Inertial Confinement Fusion Program'' under Award Number DE-NA0004144 and U.S. National Science Foundation PHY Grants No. 2205521 and No. 2020249.

This research used resources of the National Energy Research Scientific Computing Center, a DOE Office of Science User Facility supported by the Office of Science of the U.S. Department of Energy under Contract No. DE-AC02-05CH11231 using NERSC awards FES-ERCAP0031649 and FES-ERCAP0037049.

AB and RR thank Tristan Guillot, Ravit Helled, and Nadine Nettelmann for helpful discussions.  The authors gratefully acknowledge the computing time made available to them on the high-performance computers \AB{Emmy} and \AB{Lise} at the NHR Centers G\"ottingen and Berlin.
\\


%

\end{document}